\documentclass[twocolumn,showpacs,preprintnumbers,amsmath,amssymb]{revtex4}
\usepackage{graphicx}% Include figure files
\usepackage{dcolumn}% Align table columns on decimal point
\usepackage{bm}% bold math

\begin{document}

\preprint{APS/123-QED}

\title{Microwave spectral analysis by means of non-resonant parametric recovery of spin-wave signals in a thin magnetic film}

\author{S. Sch\"afer,$^1$ A.V. Chumak,$^{1,2}$ A.A. Serga,$^1$ G.A. Melkov,$^2$ and B. Hillebrands$^1$}
 %\affiliation[Also at ]{Fachbereich Physik and Forschungsschwerpunkt MINAS, Technische Universit\"at Kaiserslautern,
 %Erwin-Schr\"odinger-Str.56, 67663 Kaiserslautern, Germany}\\
\affiliation{%
  $^1$Fachbereich Physik and FSP MINAS, Technische Universit\"at Kaiserslautern, 67663 Kaiserslautern, Germany\\
  $^2$Departement of Radiophysics, Taras Shevchenko National University of Kiev, Kiev, Ukraine\\
}%

%\affiliation{

\date{\today}% It is always \today, today,

\begin{abstract}
We report on the storage and non-resonant parametric recovery of microwave signals carried by a dipolar
surface spin-wave pulse in a thin ferrimagnetic film. The information about the intensity of the spectral
components of the signal within a narrow frequency band is saved due to the excitation of a dipolar-exchange
standing spin-wave mode across the film thickness and is afterwards restored by means of parametric
amplification of this mode. The intensity of the restored signal measured for varying shifts between
the signal carrier frequency and half of the pumping frequency, which is equal to the frequency of the
standing mode, reveals information about the entire frequency spectrum of the input microwave
signal.

\end{abstract}

\maketitle

%\section{}

In our recent experiments \cite{PRL2001, PRL} a microwave signal carried by a packet of dipolar spin
waves propagating in a ferrimagnetic film was stored in the form of spatially localized spin-wave
excitations and restored thereafter by means of amplification of these excitations by parametric
pumping. For those experiments, parametrical interaction was performed by applying a microwave pumping
with a frequency $\nu_\mathrm{p}=2\cdot \nu_\mathrm{s}$ which is twice the carrier frequency
$\nu_\mathrm{s}$ of the input microwave signal.
%The given ratio between the frequencies corresponds to the pure case of parametric resonance.
Here we report on the non-resonant case of parametric restoration
with $\nu_\mathrm{p}\neq 2\cdot \nu_\mathrm{s}$. This technique enables access to the spectral
characteristics of the processes that underly both the storage and the restoration phenomena. Therefore
these studies are relevant for the basic understanding of interactions between different groups of spin
waves and especially the influence of the thermal magnon bath on the parametric interactions as well as
for technical applications, using the ability to store, retrieve and process spectral information of
microwave signals.
\begin{figure}
\includegraphics[width=0.85\columnwidth]{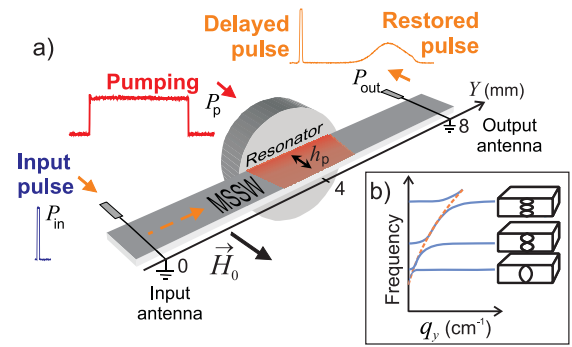}% Here is how to import EPS art
\caption{\label{setup} (Color online) a) Experimental setup of experiments reported on, consisting of a
single crystal YIG film, input and output antennae and a dielectric resonator for the application of the
pumping microwave field. The waveforms schematically depict the form of the input, pumping and restored
pulse as can be seen on the oscilloscope. b) Section of the spin-wave dispersion spectrum for a thin
magnetic film schematically showing the hybridization of MSSW (dashed line) with SSW modes (cont.
lines).}
\end{figure}

The experimental setup includes a long and narrow ($30 \times 1.2$~mm$^2$) spin-wave waveguide cut out
of a thin (5.94~$\mu$m) single crystal YIG film which is placed upon both input and output microstrip
antennae (see Fig.~\ref{setup}). The signal microwave pulse excites a packet of spin waves which
propagates along the waveguide from the input to the output antenna. The signal picked up by the output
antenna is observed with an oscilloscope after amplification and detection. The long axis of the YIG
film waveguide and therefore the propagation direction of the spin wave is perpendicular to the static
magnetic field of $H_\mathrm{0} = 1780.5$~Oe applied in the film plane. Thus the experimental geometry
generally corresponds to the case of dipolar-dominated magnetostatic surface spin waves
(MSSW)~\cite{Damon_Eshbach}. In addition, due to the finite thickness of the magnetic medium, discrete
exchange-dominated standing spin-wave modes (SSW) perpendicular to the film's surface
exist~\cite{Kalinikos}. In crossing regions of SSW and MSSW modes, both magnon groups are interacting
and a hybridization takes place, leading to a dispersion spectrum as shown in the inset of
Fig.~\ref{setup}. Thus, an excitation of SSW modes by a bypassing MSSW packet occurs. Due to the almost
negligible group velocity of SSW modes, information encoded in the traveling wave is
locally conserved in the magnetic film and therefore can be retrieved by means of parametric amplification
of the SSW even after the traveling spin-wave pulse has left the area between the antennae.

In order to amplify the SSW modes we used the parallel pumping method~\cite{Gurevich}. The pumping
magnetic field, which is parallel to the bias magnetic field, is created by the dielectric resonator
attached to the waveguide in the middle between the antennae (see Fig.~\ref{setup}). The resonator is
excited by a microwave pulse at a fixed carrier frequency of $\nu_\mathrm{p} =14.258$~GHz. This
frequency was chosen to match the doubled frequency of one SSW mode in order to obtain the
maximum efficiency of its parametric amplification. In order to avoid a direct amplification of the
traveling MSSW the pumping pulse is only applied after the spin-wave packet is detected at the output
antenna and therefore has left the area of parametric interaction. In the experiments reported on, the
pumping pulse with a power of 4~W had a duration of 7~$\mu$s and was delayed by 400~ns with respect to
the input pulse. As a result of the pumping an additional, so-called "restored", MSSW pulse was received
by the output antenna long after the original delayed MSSW pulse as is indicated in the
waveform insets of Fig.~\ref{setup}. In general the intensity, delay time and width of the restored
pulse depend on the intensity and delay time of the pumping as well as on the frequency and intensity of
the input signal pulse.
\begin{figure}
\includegraphics[width=0.75\columnwidth]{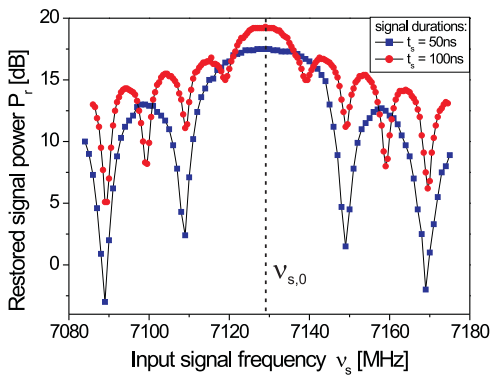}% Here is how to import EPS art
\caption{\label{results50_100} (Color online) Experimental results. Power of the restored pulse is
depicted against the carrier frequency $\nu_\mathrm{s}$ of the input signal pulse. The dots represent
the further discussed measurement with signal pulse length of $\tau_\mathrm{s} =100$~ns. As comparison,
the squares show the results for a pulse duration of $\tau_\mathrm{s} =50$~ns, revealing a broader
spectral distribution as is expected from equ.~(\ref{nrpr_linear_spec}).}
\end{figure}
In Fig.~\ref{results50_100} our experimental results on the intensity of the restored pulse vs. the
carrier frequency of the input signal are presented. The measurements were performed for the input
microwave pulse durations of $\tau_\mathrm{s} =50$~ns and $\tau_\mathrm{s} =100$~ns. The results were
obtained by keeping the output voltage of the microwave detector constant while varying the attenuation
of the signal before it enters the low noise microwave amplifier and semiconductor detector. This was
done in order to operate all relevant microwave components and the detector in a linear regime of
operation and exclude all possible nonlinear influences of the experimental setup. One can see the
dependence of the restored pulse intensity on the carrier frequency $\nu_\mathrm{s}$ of the input signal
pulse. The global maximum at $\nu_\mathrm{s, 0} =7.129$~GHz corresponds to half the pumping frequency as
expected for the pure resonant case when the carrier frequency of the MSSW packet matches the
frequency of the parametrically amplified SSW mode. The positions of auxiliary maxima (as well as minima)
are directly correlated with the Fourier spectra of the input signals. For example, the doubling of the
frequency width for a pulse with half the duration is clearly visible. The reason for that correlation
is the fact that only the intensity of spectral components of the signal pulse within the frequency
width of the parametrically amplified SSW mode is relevant for the recovery process and determines thus the intensity of the
restored pulse. As a result, the frequency resolution of the process depends on the frequency width of
the SSW mode, which our case is smaller than 1~MHz.

The absolute difference of the peak intensities for different signal pulse durations is caused by two
effects. First of all, a longer MSSW pulse excites the SSW modes more effectively due to the longer
interaction time. Second, since the delay was measured with respect to the signal pulse rising edge, the
time interval between the signal and the pumping pulses is smaller for the longer signal pulse, giving
the SSW modes less time to loose energy before the pumping sets in.

\begin{figure}
\includegraphics[width=0.75\columnwidth]{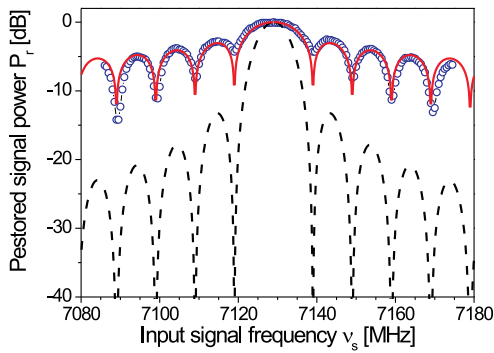}% Here is how to import EPS art
\caption{\label{all3curves} (Color online) Arrangement of experimental results for a pulse duration of
$\tau_\mathrm{s} = 100$~ns from figure~\ref{results50_100} (circles) and theoretical curve according to
equation~(\ref{nrpr_linear_spec}) (dashed line) for the input pulse signal power against the signal
frequency. The continuous line shows the restored pulse power expected from
equation~\ref{nrpr_P_restored} with $\delta = 0.77$.}
\end{figure}

In order to explain the experimental results in detail we now concentrate on the data obtained
with a pulse duration of $\tau_\mathrm{s} =100$~ns. In Fig.~\ref{all3curves} the normalized
experimental values (circular dots) are presented together with the Fourier spectrum of the input signal
pulse (dashed line)
\begin{equation} \label{nrpr_linear_spec}
P_\mathrm{s} \propto \left[ \sin{\left( \left(2\pi\nu_\mathrm{s}-2\pi\nu_\mathrm{s, 0}\right)\cdot
\dfrac{\tau_\mathrm{s}}{2}\right)} / \left(2\pi\nu_\mathrm{s}-2\pi\nu_\mathrm{s, 0} \right) \right]^2
\end{equation}
normalized with respect to the experimental values. In spite of clear similarities between these spectra
one can see that the relative intensities of the spectral components of the restored pulse are
significantly different from the frequency spectra of the input pulse, especially the range of output
power is drastically reduced for the experimental values obtained from the restored pulse. Thus no
linear connection between the intensities of the restored and input signals exists.

We interpret this deviation using a simple model based upon a general theory of parametric interaction
of spin waves~\cite{Lvov}. Two aspects are of particular interest for the explanation of our
experiments. First, the hybridization of MSSW and SSW modes and the creation of so-called
dipole-exchange gaps reported on in~\cite{PRL} is the fundamental precondition for the transfer of
energy and information from the traveling spin wave to the standing modes and back again, which enables
us to store the signal pulse information for some time. The second important aspect is the process of
parametric amplification which is responsible for the restoration of the stored signal. During this
process, two effects are important: the amplification of the standing spin wave and the amplification of
other spin-wave modes from the thermal floor. A microwave magnetic pumping field $h_\mathrm p$
parametrically generates and amplifies spin waves from the thermal floor as well as the SSW modes
pre-excited by the signal spin-wave pulse, increasing their amplitudes with time as exp($h_\mathrm p
V_\mathrm k -\Gamma_\mathrm k)t$ and exp($h_\mathrm p V_\mathrm s -\Gamma_\mathrm s)t$,
respectively~\cite{Gurevich}. The parameters $V_\mathrm{k}$, $V_\mathrm{s}$, $\Gamma_\mathrm{k}$ and
$\Gamma_\mathrm{s}$ denote the coefficients of parametric coupling with the pumping field and the
relaxation frequencies for the thermal magnons and the standing spin waves, respectively.

According to L'vov~\cite{Lvov} however, due to the competition between the magnon groups only one
dominating spin-wave group with the maximum gain factor of parametric amplification is finally excited.
This process can be described by the concept of an internal pumping field generated by the dominating
magnon group and compensating the external pumping field acting on the other magnon groups
(see~\cite{PRL, Lvov}). In our case the amplification factor $h_\mathrm p V_\mathrm k -\Gamma_\mathrm k$
of exchange spin waves with $k \approx 10^{5}$~rad/cm is always higher than the amplification $h_\mathrm
p V_\mathrm s -\Gamma_\mathrm s$ of the SSW \cite{PRL, Gurevich}. Therefore, no generation of standing
spin waves in the dipole-exchange gaps and consequently no restored MSSW pulse is observed  as long as
no input signal pulse is applied to the system.

When a signal MSSW is exciting standing spin waves however, the pumping field amplifies the standing
spin waves from the level $A_{s,0}$ which is significantly higher then the initial amplitude of the
dominating group determined by the thermal level $A_\mathrm T$. Thus during some transition times the
suppressing influence of the dominating group remains small and amplification of the SSW modes is
effective. As soon as the amplitude of those dominating spin waves exceeds a critical threshold, the
amplification of the standing spin waves and therefore the amplitude of the recovered pulse starts to be
suppressed. This moment is associated with the maximum of the recovered pulse, having an amplitude
\begin{equation}
\label{nrpr_amplitude}A_\mathrm{s, max} = A_\mathrm{s, 0}\cdot \left( \frac{A_\mathrm{k, cr}}{A_\mathrm{T}}\right)^{\delta}~,
\end{equation}
where $A_\mathrm{k, cr}$ is a critical amplitude of the thermal magnons causing the increased damping
and $\delta$ is a parameter defined by
\begin{equation}
\label{nrpr_delta}\delta = \frac{h_\mathrm p V_\mathrm{s}-\Gamma_\mathrm{s}}{h_\mathrm p
V_\mathrm{k}-\Gamma_\mathrm{k}}~.
\end{equation}
This relation describes a saturation behavior and decay for parametrically amplified SSW due to the
competition with thermal magnons. However, for the case of the constant thermal level the described
model results in a linear relation between the intensities of the restored and input signals. In order to
interpret the experimental data we supposed that the MSSW pulse irradiated by the input antenna is able
to significantly increase the level of thermal spin waves within its spectral width due to two-magnon
scattering processes~\cite{Gurevich}. We can therefore assume that the level of the thermal magnons is
increased by the signal pulse according to
\begin{equation}
\label{nrpr_thermal} A_\mathrm{T} = A_\mathrm{T, 0} + \beta \cdot A_\mathrm{s, 0} \approx \beta \cdot A_\mathrm{s, 0}
\end{equation}
assuming $\beta \cdot A_\mathrm{s, 0} \gg A_\mathrm{T, 0}$, where $A_\mathrm{T, 0}$ is the thermal
magnon level before application of the input signal. The coefficient $\beta$ describes the
influence of the signal spin wave on the thermal magnon gas. Since $A_\mathrm{s, 0} = \sqrt{P_\mathrm{s,
0}}$, the power of the restored pulse can be depicted as
\begin{equation} \label{nrpr_P_restored} P_\mathrm{r} = c \cdot P_\mathrm{s, 0}^{1-\delta}~, \end{equation}
with the constant $c = ( A_\mathrm{k, cr}/\beta )^{2 \delta} $.

A linear regress of the restored signal power against the input signal power reveals a value of 0.77 for
the parameter $\delta$ in Eq.~(\ref{nrpr_P_restored}). A look at the continuous line in
Fig.~\ref{all3curves}, representing the input power (dashed line) multiplied with $\alpha = 1 - \delta
= 0.23$ reveals the excellent accordance between the experimental data (circles) and our model, which
predicts a linear dependency between input and restored signal intensities within the logarithmic scale.
The remarkable feature visible from that curve is the drastic increase in dynamic range of a device
based upon the presented effects. The accessible range of microwave power detectable will be increased
by the coefficient $\alpha$ using the logarithmic dB-scale.

In conclusion, the non-resonant restoration of microwave signals by means of parametric pumping was
observed, basing on the storage of spin-wave information in the hybridized spin-wave spectra of a thin
YIG film. This is a first step of a deeper understanding of the process of restoration of spin-wave
pulses by parametric interaction. The simple phenomenological theoretical model proposed describes the experimental results very well. Furthermore, we propose to use the described process for a spectrum analyzer, taking advantage of the range of the output power which is compressed by $P_\mathrm{r} = c \cdot P_\mathrm{s, 0}^{\alpha}$ and thus increases the accessible dynamic range as this is usually limited by the restricted range of the analyzers' detecting component. Financial Support by the DFG within the SFB/TRR 49 is gratefully acknowledged.

%\newpage %Just because of unusual number of tables stacked at end

%\textit{Parametrically stimulated recovery of microwave signal stored in standing spin-wave modes of a magnetic film},
%\textit{Magnetostatic modes of a ferromagnet slab},
%\textit{Magnetization oszillations and waves},
%\textit{Theory of dipole-exchange spin wave spectrum for ferromagnetic films with mixed exchange boundary conditions},

\end{document}